# Filamentation of Mid-IR pulses in ambient air in the vicinity of molecular resonances

V. Shumakova,[1] S.Ališauskas,[1] P. Malevich[1], C. Gollner,[1] A. Baltuška,[1,2]
D. Kartashov,[3] A.M. Zheltikov,[4,5,6] A.V. Mitrofanov,[4,5] A.A. Voronin,[4,5]
D.A. Sidorov-Biryukov,[4,5] A. Pugžlys[1,2]*

[1]Photonics Institute, TU Wien, Gusshausstrasse 27-387, A-1040 Vienna, Austria
[2]Center for Physical Sciences & Technology, Savanoriu Ave. 231 LT-02300 Vilnius, Lithuania.
[3]Friedrich-Schiller University Jena, Max-Wien Platz 1, 07743 Jena, Germany
[4]Physics Department, M.V. Lomonosov Moscow State University, 119992 Moscow, Russia
[5]Russian Quantum Center, ul. Novaya 100, Skolkovo, Moscow Region, 143025 Russia
[6]Department of Physics and Astronomy, Texas A&M University, College Station TX, 77843-4242, USA
*Corresponding author: pugzlys@tuwien.ac.at

**Properties of filaments ignited by multi-millijoule, 90-fs mid-IR pulses centered at 3.9 µm are examined experimentally by monitoring plasma density and losses as well as spectral dynamics and beam profile evolution at different focusing strengths. By softening the focusing from strong (f=0.25 m) to loose (f=7 m) we observe a shift from plasma assisted filamentation to filaments with low plasma density. In the latter case, filamentation manifests itself by beam self-symmetrization and spatial self-channeling. Spectral dynamics in the case of loose focusing is dominated by the non-linear Raman frequency downshift, which leads to the overlap with the $CO_2$ resonance in the vicinity of 4.2 µm. The dynamic $CO_2$ absorption in the case of 3.9-µm filaments with their low plasma content is the main mechanism of energy losses and either alone or together with other nonlinear processes contributes to the arrest of intensity.**

http://dx.doi.org/10.1364/OL.99.099999

Until recent years, filamentation of femtosecond mid-infrared (mid-IR) pulses in ambient air remained a long-standing challenge because of the lack of high peak-power sources capable of exceeding critical power of self-focusing which scales quadratically with the wavelength [1]. The critical power of self-focusing for femtosecond 800 nm driving pulses in ambient air, in dependence on the pulse duration, was evaluated to be ~5-10 GW [2], which, when applying $\lambda^2$-scaling law, results in 125-250 GW for 3.9-µm pulses. Following a steady progress in the development of mid-IR optical parametric chirped pulse amplification (OPCPA) technology [3], first successful demonstration of mid-IR filaments in ambient air was recently reported [4, 5] initiating a vigorous debate in the community on the filamentation mechanisms in mid-IR. Recent numerical studies [6, 7] reveal that substantially lower ionization rates cause significantly smaller electron plasma densities in mid-IR filaments as compared to the case of filaments generated by more common 800-nm and 1030-nm near-IR drives. Next to lower plasma density, other factors, such as, saturation of higher order Kerr nonlinearity (HOKE) terms [8], shock driven walk-off of generated harmonics [7] or thermal gas density gradients [9, 10] might play a non-negligible role in filamentation of mid-IR pulses. Furthermore, through numerical simulations [6, 11] it was demonstrated that since mid-IR spectral range contains numerous vibrational resonances, energy losses due to absorption of the spectral components generated in a filament as a result of spectral broadening, needs to be considered next to ionization, rotational Raman excitation and plasma absorption [12, 13]. However in [6] and [11] self-phase modulation (SPM) was considered as the main mechanism of spectral broadening, whereas stimulated Raman scattering (SRS) was not included into the models. Here, by means of extensive experimental studies of filamentation of 3.9-µm pulses in air, we identify that namely SRS is responsible for the enhancement of dynamic absorption by $CO_2$ molecules and therefore, plays a key role in governing energy losses during filamentation. In addition, as filamentation is usually assisted by external focusing, here we show that the focusing strength strongly affects plasma density inside of mid-IR filament and discuss its' influence on the spectral dynamics and related energy losses.

Experiments were performed with a hybrid OPA/OPCPA system described in detail in [3]. In order to reach peak powers exceeding the critical power of self-focusing in mid-IR, the system was upgraded by installing an additional OPCPA stage non-collinearly pumped by 700 mJ, 70 ps pulses originating from a Nd:YAG amplifier (Fig.1). In this additional OPCPA stage, based on a Potassium titanyl arsenate (KTA) nonlinear optical crystal, idler 3.9-μm pulses are amplified to an energy exceeding 45 mJ, which is reduced to 30 mJ during compression in a grating compressor with the transmission efficiency of 66%. As pulses are compressed to a duration of 90 fs peak powers higher than 330 GW are achieved.

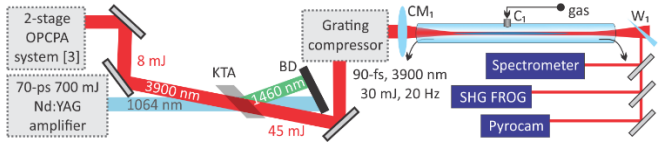

Figure 1. Layout of the upgraded OPA/OPCPA system and the experimental arrangement with an open-ended gas tube; BD – beam dump, KTA – Brewster-angle cut nonlinear optical crystal. $CM_1$ – spherical mirror, $C_1$ – open-ended gas cell, $W_1$ – $CaF_2$ wedge.

Filaments in ambient air were generated by focusing 3.9-μm pulses with spherical mirrors having focal lengths ranging from 0.25 m to 7 m. For the characterization of generated filaments we monitored the energy transmitted through the filament, ionic plasma density, side plasma luminescence, spectra and temporal profiles of pulses, as well as a variation of the beam profile along the filament. The spectra were recorded using a scanning monochromator (Digikrom, CVI) and a liquid nitrogen cooled InSb photodetector. Temporal pulse profiles were determined by second harmonic generation frequency resolved optical gating (SHG FROG) apparatus and spatial beam profiles were detected with a pyroelectric beam profiling camera (Pyrocam III). Relative density of ionic plasma generated as a consequence of ionization during filamentation was measured with a plasma capacity probe [14], consisting of two flat 1x1 $cm^2$ copper electrodes separated by a distance of 1 cm and charged to 2 kV DC voltage. A signal governed by the current produced by $O_2^-$ and $O_2^+$ ions gives information about the plasma density on a microsecond time scale [15]. In addition, side plasma luminescence was recorded with a digital photo camera (Canon 350D). The luminescence is caused mainly by neutral (excited by an electron impact) and by ionized $N_2$ molecules and takes place on a nanosecond time scale, i.e., before the plasma channel expands, which allows us to monitor both, the diameter and the length of the filament.

Spectral and structural transformations during filamentation assisted by external focusing were found to be strongly dependent on the focusing strength.

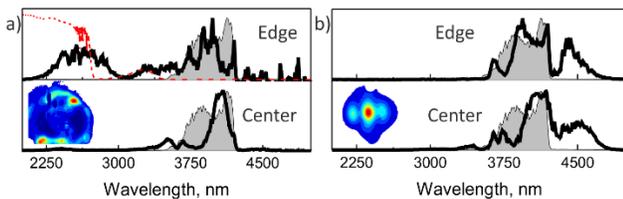

**Figure 2.** Spectra (thick black line) and beam profiles recorded after the filament in the cases of focusing with spherical mirrors of f=0.25 m (a), leading to the formation of plasma, and of f=7 m (b), when negligible ionic plasma content was detected; gray shaded area represent the original spectra; input pulse energy in both cases was 30 mJ; the spectra were recorded in the center and at the edge of the beam; red dashed line in the top panel (a) represents the transmission spectrum of a 1-cm thick BK7 window used to block the mid-IR part of the spectrum.

When 30 mJ pulses are hard-focused with spherical mirrors of f≤0.75 m, a donut-like beam profile is observed after the filament. In this case, the filament is evidenced by a bright plasma channel. Spectra recorded in the periphery of the donut-shaped beam feature strong plasma blue-shift, with the blue wing of the spectrum reaching 2 μm wavelength (top panel of Fig. 2a). In this case, around 30% of the energy is transferred into wavelengths below 3 μm as determined by blocking the mid-IR radiation by a 1-cm thick BK7 substrate. The spectrum recorded in the central part of the beam (bottom panel of Fig. 2a) extends from 3.5 to 4.1 μm, i.e., it does not exhibit any substantial spectral broadening, which reveals that the donut-shape beam profile is caused by plasma refraction. The steady spectrum in the center of the beam corresponds to the leading front of the pulse, wherein the intensity, necessary for plasma generation, is not yet reached. By increasing the focal length (0.75 m<f<2.5 m) both, ionic plasma manifestation and blue-sided spectral broadening become less pronounced and the beam retains its spatial quality at increasingly higher pulse energies. In the case of loose focusing (f=7 m), when plasma luminescence is not detectable, the spectral dynamics is similar in the center and at the periphery of the beam. They are dominated by a red shift of the spectrum (Fig. 2b), typical for the case of molecular gases and attributed to SRS [16]. A sharp drop of intensity in the vicinity of 4.2 μm is linked to the absorption by $CO_2$.

In the case of loose focusing, a rather weak plasma generation, evidenced by vanishing electrical signal and the absence of visible luminescence, was observed. In order to confirm filamentation of the mid-IR pulses under these circumstances, we examined the evolution of the beam profile along the propagation direction (Fig. 3a, b). The beam profile evolution in the vertical (y) and horizontal (x) directions was recorded for compressed pulses, when self-focusing due to Kerr lensing is pronounced and filamentation takes place, and for attenuated long chirped pulses (stretched to >3 ps pulse duration by detuning the OPCPA compressor), which corresponds to the case of linear propagation without noticeable spectral broadening. In the case of stretched pulses the measured beam radii correspond well to the calculated radii evolution of a Gaussian beam (Fig. 3a,b). Note that the initial beam profile is slightly elliptical and the beam is astigmatic, exhibiting different divergences in (x) and (y) directions. By contrast, when compressed pulses are focused by a f=7 m mirror, beam radii rapidly shrink to sub-2 mm and remain at this level for several meters. In the case of compressed pulses, the minimum beam radius of ~0.4 mm at the $1/e^2$ level was measured along both transverse axes at the distance of ~550 cm from the focusing mirror. As it can be seen from Fig. 3c, self-symmetrization of the beam profile takes place already at the position of the focusing mirror and is sustained for more than 5 meters. At the distance of ~6 m, two distinct side-lobes emerge along the long axis of the ellipse representing the initial beam (right most profile in Fig. 3e). In contrast to multiple filamentation caused by modulation instability, this beam profile evolution is reproducible from shot to shot and is presumably caused by the nucleation, in the direction of initial ellipticity, of low intensity structure distributed around the filament core [17]. The observed beam self-symmetrization in the case of 30-mJ, 90-fs pulses reveals that an arrest of the intensity takes place, which is a robust feature of filamentation.

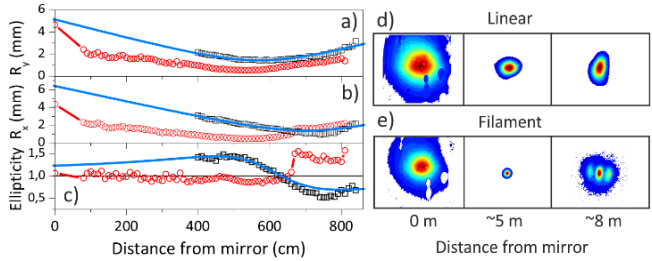

**Figure 3.** a) and b) –evolution of the radii of the beam in y- and x-directions at $1/e^2$ level along the propagation of compressed (open red dots) and stretched to >3 ps (open black squares) pulses, the solid blue lines represent calculated beam diameter evolution in the case of linear focusing with a f=7 m spherical mirror; c) dynamics of ellipticity of the beam along the propagation of compressed (open red dots) and stretched to >3 ps (open black squares) pulses; the solid blue line represents the ellipticity corresponding to the calculated curves presented in panels a) and b); d) and e) - beam profiles measured accordingly (from left to right) at the position of focusing mirror, minimal beam-waist position and at the distance of 8 m from the focusing mirror for stretched (d) and compressed (e) pulses.

Capacity probe measurements (Fig. 4a) reveal that with the softening of focusing from f=0.25 m to f=7 m, both the current between the electrodes, resulting in a voltage drop on a resistor, and the plasma luminescence decrease by a factor of more than 20. The fact that the measured peak luminescence signal and spatially-integrated luminescence (not shown) behave similarly, reveals that both plasma density and the total amount of ions generated in a filament decrease proportionally with softening the focusing. Note, that in the case of f=7 m plasma luminescence was not detectable with a photo camera.

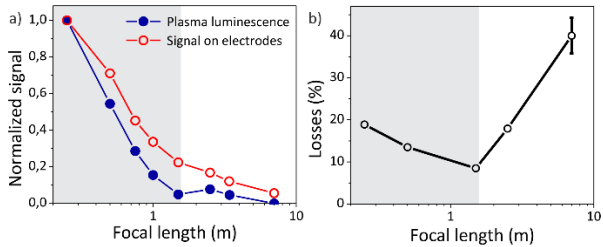

**Figure 4** (a) Dependence of the maximum signal on the electrodes of the plasma capacity probe (open circles) and luminescence intensity (solid circles) on the strength of focusing; (b) dependence of the energy losses on the strength of focusing; shadowed area corresponds to the focusing range at which losses decrease with softening of the focusing.

By contrast, the energy losses during filamentation, with respect to the focusing behave quite different (Fig. 4b.). As focusing softens from f=0.25 m to f=1.5 m the losses decrease from 17% to <10%, which follows the trend of decreasing plasma signature. With further softening of the focusing towards f=7 m, in which case the plasma density is rather low, the losses surge to the level of >35%. This observed filamentation feature of mid-IR pulses is remarkably different from the well-studied near-IR scenario. In the case of filaments generated with intense near-IR (800-nm) pulses, up to 60% of the initial pulse energy can be lost mainly due to plasma generation and absorption [12], which becomes weaker with softer focusing. Although the amount of deposited energy in this case depends on a number of parameters, including the focusing strength, the losses are mainly caused by ionization of air, which, we believe is the same case for strongly focused 3.9-μm pulses. Lower absolute values of losses (17% instead of 60%) can be attributed to multiple-filamentation reported in [12] and lower ionization rates in the case of mid-IR pulses [18].

To examine the losses in the case of 7-m focusing, we generated filaments in the main constituent parts of atmospheric air, namely $N_2$, $O_2$ and in the reference atomic gas Ar and monitored spectra and ionic plasma signals. The compressor of our OPCPA system in each case was adjusted for the highest filamentation losses. In order to avoid additional nonlinear effects in the cell windows, the experiments were performed in a 4-m long tube with open ends (Fig.1) as gases under investigation were flown through the tube under slight overpressure.

The spectral broadening in pure $O_2$ and $N_2$ was found to be more efficient than in ambient air and is dominated by a red shift due to stimulated Raman scattering [16]. In Ar spectral broadening in the mid-IR is more symmetric and is caused mainly by SPM, without any significant plasma blue-shift which was reported in the case of focusing 12 mJ pulses with a f=2 m spherical mirror into a gas cell with Ar at 4.5 bar [19].

| Gas | $I_p$ (eV) | C (%) | V (mV) | L (%) |
|-----|------------|---------|--------|-------|
| Air |            | 100     | 17     | 36    |
| $N_2$ | 15.6     | 78,084  | 10     | 8     |
| $O_2$ | 12.08    | 20,9476 | 42     | 13    |
| Ar  | 15.76      | 0,934   | 115    | 2     |

**Table 1.** Ionization potential ($I_p$), concentration in air (C), ionic plasma signal (V), and losses measured after the filaments ignited in the main constitutive components of air. Spectra measured after the open-ended tube are presented in the last column. In the top-right column OPCPA output spectrum and $CO_2$ absorption spectrum are shown.

As seen from Table 1, where the results are summarized, measurements with the capacity probe reveal weak ionic plasma generation in $N_2$ (~10 mV signal) and four times higher in $O_2$ (~42 mV signal), which is explainable by the lower ionization potential $I_p$ of oxygen. In air, the the signal of the plasma capacity probe is determined by an even more complex plasma chemistry than in the cases of pure $N_2$ or $O_2$. The signal in air is 17 mV, more similar in magnitude to $N_2$. However, the energy losses, 8% in the case of $N_2$ and 13% in the case of $O_2$, are significantly lower than that in air (>36%). The losses do not correlate with the maximum of the capacity probe signal, suggesting that they are not related to the generation of weak plasma. Furthermore, in Ar, which has an ionization potential similar to that of $N_2$, capacity probe measurements provide a more than 10 times larger signal as compared to the case of $N_2$ (explainable by a substantially longer plasma lifetime in Ar [20]). At the same time, filamentation losses in the case of Ar are hardly detectable. Those observations suggest that neither plasma generation nor rotational Raman excitation of $N_2$ and $O_2$ [13] are responsible for the observed >36% losses during filamentation in air when filamentation is assisted by loose focusing.

In order to determine the origin of the losses we examined spectral, temporal and spatial transformations of 3.9 μm pulses during filamentation (Fig. 5). The obtained data allow us to retrieve intensity evolution along the filament and to characterize filamentation dynamics in the case of loose focusing. Filamentation dynamics was found to be very sensitive to the chirp of propagating mid-IR pulses, which is subject to a separate investigation. For these particular measurements, the compressor of the OPCPA system was tuned to maximize the losses after the filament. As determined by SHG FROG characterization, the optimal input pulses were slightly positively pre-chirped to the duration of 130 fs FWHM.

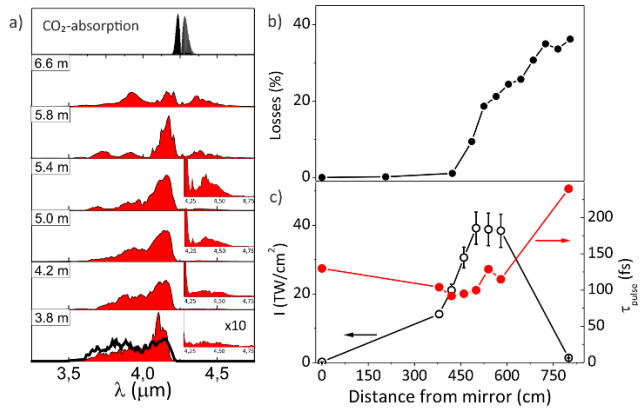

**Figure 5.** Measured evolution of the spectral shape (a), losses (b), pulse duration and intensity (c) during filamentation of 30-mJ pulses assisted by 7-m focusing; in (a) the distance from the focusing mirror is indicated in the panels; the area under the spectra are scaled to the pulse energy; the initial spectrum is shown by a black line in the lower panel; in the insets the red part of the spectrum is zoomed with x10 magnification.

As seen from Fig. 5a, during propagation through the first 4 m of air the spectrum is only slightly transformed by a redistribution of the spectral content from the blue to the red wing (lower panel in Fig. 5a) which is a recognizable feature of SRS [16]. Also, due to the anomalous dispersion of air in the spectral range of 3.6-4.2 µm [21] during this initial propagation pulses are recompressed to 90 fs (Fig. 5c). Small (<3%) losses during the first four meters of propagation (Fig. 5b) might be caused both, by absorption of $CO_2$ and rotational Raman excitation of $N_2$ and $O_2$ molecules [13].

With further propagation the intensity sharply rises (Fig. 5c) which is a combined effect of pulse compression and reduction of the beam diameter (Fig. 3a, b). The sharp rise of the intensity is followed by a pronounced red shift of the spectrum, which correlates with the sharp increase of the losses. The absorption coefficient of air at the wavelength of 4.26 µm, corresponding to the $CO_2$ resonance is $10^{-2}$ cm$^{-1}$ [22], which means that nearly 2/3 of 4.26-µm light is absorbed after traveling 1 m of air. We observe intensing stabilization of the 40 TW/cm$^2$ level between 5 and 6 m of propagation, which is explained by a continued decrease of the beam diameter (Fig. 3a, b) and still growing losses. Around meter 6 of the propagation after the focusing mirror a noticeable spectral change takes place as the pulses undergo temporal splitting, which makes evaluation of the intensity complicated.

As seen in Fig. 5a, for propagation distances over 5 m, spectral broadening is still present to the red of the $CO_2$ absorption band. We argue, that SRS-driven continuous red-shift cannot, at least on its own, reemerge beyond the $CO_2$ resonance. Therefore, a different wave-mixing mechanism tolerant to bandwidth interruption, such as SPM, must be responsible for relaunching spectral broadening in the red wing.

In conclusion, filamentation of mid-IR pulses in ambient air is strongly affected by focusing, but in a rather different way than it is known for near-IR filamentation. Experimental results suggest that the physical origin of losses and the mechanism of stabilization of the spatial collapse in mid-IR filaments strongly depends on the focusing conditions. In the case of strong focusing, the generated plasma plays a dominant role in determining spectral dynamics and losses during filamentation. In the case of loose focusing filamentation can be achieved with low plasma density. Under these conditions, although the concentration of $CO_2$ in ambient air is in the order of 500 ppm, this atmospheric constituent plays a dominant role in determining losses during filamentation, as spectral amplitude, continuously red-shifted by rotational SRS, provides steady supply for resonant $CO_2$ absorption. Consequently, the dynamic $CO_2$ absorption plays an important role in filamentation of 3.9-µm pulses and either by itself or in combination with other (nonlinear) processes contribute to the arrest of the intensity and to the prevention of beam collapse.

**Funding.** Austrian Science Fund (FWF) (P 26658, P 27577);